\documentclass[aps, prb, twocolumn, superscriptaddress, showpacs]{revtex4}
\usepackage{mathrsfs, amsfonts, amsmath, amssymb, graphicx, bm, epsfig}
\usepackage[colorlinks, linkcolor=blue, anchorcolor=blue, citecolor=red]{hyperref}

\newcommand\B{\rule[-1.0ex]{0pt}{0pt}}

\begin{document}
\title{Adiabatic Electron Dynamics in Antiferromagnetic Texture}
\author{Ran Cheng}
\email{rancheng@physics.utexas.edu}
\affiliation{Department of Physics, University of Texas at Austin, Austin, Texas 78712, USA}
\author{Qian Niu}
\affiliation{Department of Physics, University of Texas at Austin, Austin, Texas 78712, USA}
\affiliation{International Center for Quantum Materials, Peking University, Beijing 100871, China}

\begin{abstract}
   Adiabatic dynamics of conduction electrons in antiferromagnetic (AFM) materials with slowly varying spin texture is developed. Quite different from the ferromagnetic (FM) case, adiabaticity in AFM texture does not imply perfect alignment of conduction electron spins with background profile, instead, it introduces an internal dynamics between degenerate bands. As a result, the orbital motion of conduction electrons becomes spin-dependent and is affected by two emergent gauge fields: one of them is the non-Abelian version of what has been discovered in FM systems; the other leads to an anomalous velocity that has no FM counterpart. Two examples with experimental predictions are provided.
\end{abstract}

\maketitle

\section{Introduction}

Interplay between current and magnetization is an essential issue underpinning the field of spintronics~\cite{ref:spintronics}, which consists of two reciprocal problems: control of current through magnetization with a known configuration, and its converse, \emph{i.e.}, control of magnetization dynamics via applied current. In ferromagnetic (FM) materials with slowly varying spin texture $\bm{m}(\bm{r},t )$ over space and time, these issues can be solved by assuming that conduction electron spins always follow the background texture profile, known as the adiabatic approximation~\cite{ref:BerryPhase, ref:Adiabaticity}. The microscopic basis underlying adiabaticity is the strong exchange coupling $H=-J\sigma\cdot\bm{m}(\bm{r},t )$ between conduction electron spins and local magnetic moments, through which spin mistracking with the background causes large energy penalty and becomes highly unfavorable~\cite{ref:Shengyuan}.

Under adiabatic approximation, the current -- magnetization interaction is recast into an emergent electrodynamics, in which its reciprocal influence boils down to a simple electromagnetic problem. Specifically, by diagonalizing the local exchange Hamiltonian via local unitary transformation, fictitious electric and magnetic fields are generated into the orbital dynamics~\cite{ref:Shengyuan, ref:Volovik, ref:SMF, ref:Karin}
\begin{align}
 E_i&=\frac12\bm{m}\cdot(\partial_t\bm{m}\times\partial_i\bm{m})=\frac12\sin\theta(\partial_t\theta\partial_i\phi-\partial_i\theta\partial_t\phi), \notag \\
 B_i&=-\frac14\varepsilon_{ijk}\bm{m}\cdot(\partial_j\bm{m}\times\partial_k\bm{m})=-\frac12\varepsilon_{ijk}\sin\theta\partial_j\theta\partial_k\phi, \notag
\end{align}
where $\theta(\bm{r},t)$ and $\phi(\bm{r},t)$ are spherical angles specifying the direction of $\bm{m}$. As a consequence, the influence of background texture is represented by an effective Lorentz force $\bm{F}=s\hbar(\bm{E}+\dot{\bm{r}}\times\bm{B})$ exerted on conduction electrons, where $s=+1 (-1)$ denotes spin-up (-down) bands. The electric and magnetic components of the Lorentz force are responsible for the spin motive force~\cite{ref:Shengyuan, ref:SMF} and the topological Hall effect,~\cite{ref:THE} respectively. In turn, back-reaction of the Lorentz force provides an interpretation to the current-induced spin torque exerted on magnetic texture.~\cite{ref:ZhangShoucheng, ref:Tserkovnyak, ref:STT} In a formal language, the adiabaticity induces an effective gauge interaction $\mathcal{L}_{\mathrm{int}}=j_\mu\mathcal{A}_\mu$, where current $j_\mu$ acquires gauge charge according to $s=\pm1$, and $\mathcal{A}_\mu=\mathcal{A}_\mu(\bm{m},\partial\bm{m})$ is the effective electromagnetic potential representing space-time dependence of the texture. Variation over the current $\delta\mathcal{L}_{\mathrm{int}}/\delta j_\mu=0$ yields the effective Lorentz force; and varying over the magnetization $\delta\mathcal{L}_{\mathrm{int}}/\delta\bm{m}=0$ produces the spin-transfer torque. Thus the current-magnetization interaction is reciprocal.

However, the above picture apparently fails in antiferromagnetic (AFM) materials where neighboring magnetic moments are antiparallel. Conduction electrons are not able to adjust spins with local moments that change orientation on atomic scale. Nevertheless, the staggered order parameter $\bm{n}=(\bm{M}_A-\bm{M}_B)/2M_s$ can be slowly varying over space-time, where $\bm{M}_A$ and $\bm{M}_B$ are the alternating local moments and $M_s$ denotes their magnitudes. A natural question is whether a slowly varying staggered order still renders adiabatic dynamics of conduction electrons in some other sense. This is desired knowledge for studying spin transport in AFM materials, especially the quest for current-magnetization interaction as that for FM materials.

In spite of recent theoretical~\cite{ref:AFMTheory} and experimental~\cite{ref:AFMExperiment} progress, this problem has never been addressed. But, at the same time, AFM materials are believed to be promising candidates for new thrusts of spintronics,~\cite{ref:AFMSpintronics} partly due to their tiny anisotropy, robustness against external magnetic perturbations, and absence of demagnetization, which brings prevailing advantages for experimental control. In this paper, we develop the effective electron dynamics in a smooth AFM texture by applying the non-Abelian Berry phase theory~\cite{ref:Dimi,ref:NABerryPhase,ref:Dalibard} on energy bands that are doubly degenerate. The physics of adiabaticity in AFM materials is found to be an internal dynamics between degenerate bands which can be attributed to a SU(2) Berry curvature. When translating into spin dynamics, the adiabaticity no more indicates spin alignment with the background, but a totally new evolution principle [Eq.~\eqref{eq:ds}]. Aside from spin dynamics, the orbital motion of conduction electrons is coupled to two different gauge fields: one leads to the non-Abelian generalization of the effective Lorentz force; the other results in an anomalous velocity that is truly new and unique to AFM systems. With comparisons to FM materials, this paper provides a general framework on how a given AFM texture affects the dynamics of conduction electrons. The other side of the story, \emph{i.e.}, back-reaction of current on the AFM texture, will appear in a forthcoming publication.

The paper is organized as follows. In Sec.~II, the general formalism is presented where iso-spin is introduced. In Sec.~III, our central results~\eqref{eq:EOMtotal} are derived, followed by discussions on non-Abelian Berry phase and monopole, spin and orbital dynamics of conduction electrons, and comparisons of AFM electron dynamics with its FM counterparts. In Sec.~IV, two examples are provided, and the paper is summarized in Sec.~V. Mathematical derivations are included in the Appendixes.

\section{Formalism}

Consider an AFM system on a bipartite lattice with local magnetic moments labeled by alternating $\bm{M}_A$ and $\bm{M}_B$. The spin of a conduction electron couples to the local moments by the exchange interaction $J(\bm{M}/M_s)\cdot\bm{\sigma}$, where $\bm{\sigma}$ denotes the spin operator of the conduction electron, and $\bm{M}$ flips sign on neighboring $A$ and $B$ sublattice sites. In spite of antiparallel of neighboring moments, the staggered order parameter $\bm{n}=(\bm{M}_A-\bm{M}_B)/2M_s$ usually varies slowly over space and time, and we can treat it as a continuous function $\bm{n}(\bm{r},t)$. Accordingly, the conduction electron is described by a nearest-neighbor tight-binding Hamiltonian locally defined around $\bm{n}(\bm{r},t)$:
\begin{equation}
 \mathcal{H}(\bm{n}(\bm{r},t))=
 \begin{bmatrix}
 -J\bm{n}\cdot\bm{\sigma} \ \ & \gamma(\bm{k})\\
 \gamma^*(\bm{k})             & J\bm{n}\cdot\bm{\sigma}
 \end{bmatrix}
\end{equation}
where $\gamma(\bm{k})=-t\sum_{\bm{\delta}}e^{i\bm{k}\cdot\bm{\delta}}$ is the hopping term with $\bm{\delta}$ connecting nearest neighboring $A-B$ sites (we set $\hbar=1$). In general $J$ can be negative, but we assume a positive $J$ throughout this paper.

\begin{figure}[t]
   \centering
   \includegraphics[width= 0.92\linewidth]{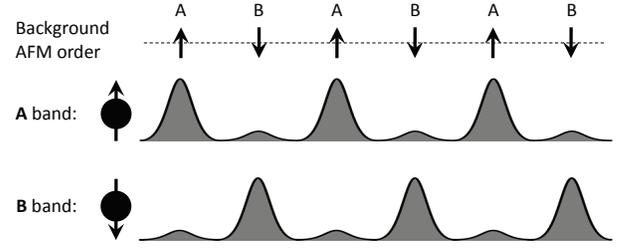}
   \caption{A schematic view of Bloch waves in the lower band. Sub-band $A$ means a local spin up electron has a larger probability on the $A$ sites and a smaller probability on the $B$ sites; sub-band $B$ means the opposite case. They are degenerate in energy and their wave functions have a finite overlap depending on the ratio of $J/\varepsilon$.}{\label{Fig:SDW}}
\end{figure}

The local band structure can be easily solved as $\pm\varepsilon (\bm{k})$ with $\varepsilon(\bm{k})=\sqrt{J^2+|\gamma(\bm{k})|^2}$, and in the adiabatic limit we neglect transitions between $\varepsilon$ and $-\varepsilon$. Each of the two bands are doubly degenerate, and without loss of generality we will focus on the lower band $-\varepsilon$ with the two sub-bands labeled by $A$ and $B$, whose wave functions are $|\psi_a\rangle=e^{i\bm{k}\cdot\bm{r}}|u_a\rangle$ and $|\psi_b\rangle=e^{i\bm{k}\cdot\bm{r}}|u_b\rangle$. The Bloch waves $|u_a\rangle=|A(\bm{k})\rangle|\!\uparrow(\bm{r},t)\rangle$ and $|u_b\rangle=|B(\bm{k})\rangle|\!\downarrow(\bm{r},t)\rangle$ maintain local periodicity around $(\bm{r},t)$, where
\begin{align}
 |\!\uparrow\!(\bm{r},t)\rangle=\!
 \begin{bmatrix}
  e^{-i\frac{\phi}2}\cos\frac{\theta}2\\
  e^{i\frac{\phi}2}\sin\frac{\theta}2
 \end{bmatrix};\
 |\!\downarrow\!(\bm{r},t)\rangle=\!
 \begin{bmatrix}
 -e^{-i\frac{\phi}2}\sin\frac{\theta}2\\
  e^{i\frac{\phi}2}\cos\frac{\theta}2
 \end{bmatrix}
 \label{eq:wavefunctions}
\end{align}
are local spin wave functions with $\theta=\theta(\bm{r},t)$ and $\phi=\phi(\bm{r},t)$ being spherical angles specifying the orientation of $\bm{n}(\bm{r},t)$. The periodic parts are spinors in the pseudo-spin space furnished by the $A-B$ sublattices,
\begin{subequations}
\begin{align}
  |A(\bm{k})\rangle=\frac{[\varepsilon(\bm{k})+J,\    |\gamma(\bm{k})|]^\mathrm{T}}{\sqrt{(\varepsilon(\bm{k})+J)^2+|\gamma(\bm{k})|^2}}, \\ |B(\bm{k})\rangle=\frac{[\varepsilon(\bm{k})-J,\    |\gamma(\bm{k})|]^\mathrm{T}}{\sqrt{(\varepsilon(\bm{k})-J)^2+|\gamma(\bm{k})|^2}},
\end{align}
\end{subequations}
which exhibit opposite spatial patterns schematically illustrated in Fig.~\ref{Fig:SDW}. While $\langle \psi_a|\psi_b\rangle=0$ due to the orthogonality of local spin eigenstates, $\langle A(\bm{k})|B(\bm{k})\rangle$ does not vanish, and we define this overlap as
\begin{equation}
  \xi(\bm{k})\!=\!\langle A(\bm{\bm{k}})|B(\bm{k})\rangle\!=\!\frac{|\gamma(\bm{k})|}{\sqrt{J^2+|\gamma(\bm{k})|^2}}\!=\!\frac{\sqrt{\varepsilon^2-J^2}}{\varepsilon}, \label{eq:overlap}
\end{equation}
which is a key parameter in our theory and $\xi<1$. It reaches maximum at the Brillouin zone (BZ) center and vanishes at the BZ boundary. From Eq.~\eqref{eq:overlap} we know $\xi(\bm{k})$ is a system parameter determined by the band structure, and it is constant since the energy conservation $\dot{\varepsilon}=0$ requires $\dot{\xi}=0$. If $J$ tends to infinity, the overlap $\xi(\bm{k})$ will vanish and the two subbands will be effectively decoupled, by which the system will become a simple combination of two independent FM subsystems.

We adopt the semiclassical approach to construct effective electron dynamics,~\cite{ref:Dimi} where an individual electron is described by a wave packet $|W\rangle=\int\mathrm{d}\bm{k}w(\bm{k}-\bm{k}_c)[c_a|\psi_a\rangle +c_b|\psi_b\rangle]$, where $\int\mathrm{d}\bm{k}\bm{k}|w(\bm{k}-\bm{k}_c)|^2=\bm{k}_c$ gives the center of mass momentum, and $\langle W|\bm{r}|W\rangle=\bm{r}_c$ is the center of mass position. The coefficients $c_a$ and $c_b$ reflect relative contributions from the two sub-bands and $|c_a|^2+|c_b|^2=1$. Since now the band is degenerate, non-Abelian formalism~\cite{ref:Adiabaticity,ref:Dimi,ref:NABerryPhase,ref:Dalibard} must be invoked (see Appendix A), where dynamics between the $A$ and $B$ subbands introduces an internal degree of freedom represented by the isospin vector
\begin{align}
\bm{\mathcal{C}}&=\{c_1, c_2, c_3\} \notag\\
&=\{2\mathrm{Re}(c_ac_b^*), -2\mathrm{Im}(c_ac_b^*), |c_a|^2-|c_b|^2\}.
\end{align}
The electron dynamics is characterized by the equations of motion of the three parameters $\bm{k}_c$, $\bm{r}_c$, and $\bm{\mathcal{C}}$, which can be obtained from variational principles with the effective Lagrangian $\mathcal{L}=\langle W|(i\frac{\partial}{\partial t}-\mathcal{H})|W\rangle$.~\cite{ref:Adiabaticity,ref:Dimi} The detailed derivations are presented in Appendix A, here we only write down the results,
\begin{subequations}
\label{eq:EOM}
 \begin{align}
  \dot{\bm{\mathcal{C}}}&=2\bm{\mathcal{C}}\times(\bm{\mathcal{A}}^r_\mu\dot{r}_\mu +\bm{\mathcal{A}}^k_\mu\dot{k}_\mu), \label{eq:EOMeta}\\
  \dot{k}_\mu& = \partial^r_\mu\varepsilon +\bm{\mathcal{C}}\cdot[\bm{\Omega}^{rr}_{\mu\nu}\dot{r}_\nu+\bm{\Omega}^{rk}_{\mu\nu}\dot{k}_\nu], \quad \label{eq:EOMk}\\
  \dot{r}_\mu& = -\partial^k_\mu\varepsilon -\bm{\mathcal{C}}\cdot[\bm{\Omega}^{kr}_{\mu\nu}\dot{r}_\nu+\bm{\Omega}^{kk}_{\mu\nu}\dot{k}_\nu], \label{eq:EOMr}
 \end{align}
\end{subequations}
where $r_\mu=(t,\bm{r}_c)$, but $k_\mu=(0,\bm{k}_c)$ has no temporal component. In Eqs.~\eqref{eq:EOM} the $\cdot$ and $\times$ denote scalar and cross products in the isospin vector space. The Berry curvatures $\bm{\Omega}$ are obtained from the gauge potentials $\bm{\mathcal{A}}$ defined on the $A$ and $B$ sub-bands, for instance,
\begin{subequations}
\begin{align}
  &[\bm{\mathcal{A}}^r_\mu\cdot\bm{\tau}]_{ij}=i\langle u_i|\partial^r_\mu|u_j\rangle \label{eq:Berrypotential}\\
  &\bm{\Omega}^{rr}_{\mu\nu}=\partial^r_\mu \bm{\mathcal{A}}^r_\nu-\partial^r_\nu\bm{\mathcal{A}}^r_\mu +2\bm{\mathcal{A}}^r_\mu\times\bm{\mathcal{A}}^r_\nu, \label{eq:rrcurv}
\end{align}
\end{subequations}
where $\bm{\tau}$ is a vector of Pauli matrices representing the isospin and $i,j$ run between $a,b$. Other components of the Berry curvatures are explained in Appendix A.

\section{Electron Dynamics}

Equipped with Eqs.~\eqref{eq:EOM}, we are now ready to derive the dynamics of an individual electron. Note that the isospin vector $\bm{\mathcal{C}}$ itself is not gauge invariant in the sense that different choice of spin wave functions results in different $\bm{\mathcal{C}}$. Therefore, in deriving electron dynamics we need to relate $\bm{\mathcal{C}}$ to the real spin defined by $\bm{s}=\langle W|\bm{\sigma}|W\rangle$ (in unit of $\frac12$) which is a physical variable and fully gauge invariant. While detailed derivations are lengthy and sophisticated, which is left for Appendix B, the final results are quite simple and elegant:
\begin{subequations}
\label{eq:EOMtotal}
 \begin{align}
  \dot{\bm{s}} &= (1-\xi^2)(\bm{s}\cdot\bm{n})\dot{\bm{n}}, \label{eq:ds}\\
  \dot{\bm{k}} &= -\frac12\bm{n}\cdot(\nabla\bm{n}\times\dot{\bm{s}}), \label{eq:dk}\\
  \dot{\bm{r}} &= -\partial_{\bm{k}}\varepsilon -\frac12(\bm{s}\times\bm{n})\cdot\dot{\bm{n}}\ \partial_{\bm{k}}\ln\xi,\qquad \label{eq:dr}
 \end{align}
\end{subequations}
where $\dot{\bm{n}}=\partial_t\bm{n}+(\dot{\bm{r}}\cdot\nabla)\bm{n}$, and we have omitted subscript $c$ of $\bm{r}_c$ and $\bm{k}_c$ for convenience of following discussions. Equations.~\eqref{eq:EOMtotal} are the fundamental equations of motion of a conduction electron in an AFM material with slowly varying texture, which are represented by joint evolutions of three variables $(\bm{s}, \bm{k}, \bm{r})$. An essential feature distinguishing the AFM electron dynamics from its FM counterpart lies in Eq.~\eqref{eq:ds}, from which we know that the real spin $\bm{s}$ does not follow the background order parameter $\bm{n}$ in the adiabatic limit.

\emph{Spin dynamics}. The motion of $\bm{s}$ can be decomposed into a superposition of two motions: one strictly follows $\bm{n}$ (for stationary $\bm{\mathcal{C}}$) and the other represents mistracking with $\bm{n}$ (for dynamical $\bm{\mathcal{C}}$), where the latter originates from dynamics between the $A$ and $B$ sub-bands and is unique to AFM materials (see examples in Sec.~IV). It is worth emphasizing that the mistracking between $\bm{s}$ and $\bm{n}$ has nothing to do with any non-adiabatic process, but is entirely due to the non-Abelian nature of the problem. The overall spin evolution can be attributed to the accumulation of a SU(2) non-Abelian Berry phase $\mathcal{P}\exp[-i\int\bm{\mathcal{A}}^r_\mu\cdot\bm{\tau}\mathrm{d}r_\mu]$ along the electron trajectory~\cite{ref:NABerryPhase}. As compared with its U(1) counterpart in FM materials~\cite{ref:Adiabaticity,ref:Shengyuan,ref:Volovik,ref:SMF,ref:Karin,ref:THE,ref:ZhangShoucheng,ref:Tserkovnyak,ref:STT}, which can be regarded as the magnetic flux of a Dirac monopole located at the center of the sphere spanned by $\bm{m}$, the SU(2) Berry phase here can be related to a 't~Hooft-Polyakov monopole in the parameter space. Detailed discussions on the monopole can be found in Appendix C, and the key point here is that a recent proposal of artificial 't~Hooft-Polyakov monopole~\cite{ref:BPSmonopole} can be realized in our AFM texture systems.

\begin{figure}[t]
   \centering
   \includegraphics[width=0.95\linewidth]{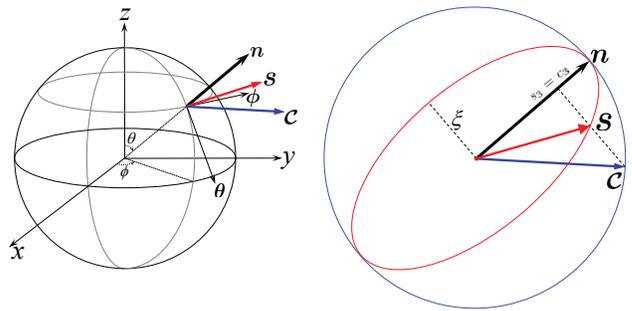}
   \caption{(Color online) Left panel: the isospin vector $\bm{\mathcal{C}}$ (blue arrow) in the local frame: $\bm{\mathcal{C}}=c_1\bm{\theta}+c_2\bm{\phi}+c_3\bm{n}$, where $\bm{\theta}$ and $\bm{\phi}$ are spherical unit vectors. Right panel: In our particular gauge, $\bm{\mathcal{C}}$ (blue) is coplanar with $\bm{n}$ and $\bm{s}$ (red). The tip of $\bm{\mathcal{C}}$ moves on a unit sphere, whereas tip of $\bm{s}$ is constrained on the ellipsoid whose semi-major axis is $\bm{n}$ and semi-minor axis having length $\xi$. \label{Fig:spinfootball}}
\end{figure}

Since $|\!\uparrow\!(\bm{r},t)\rangle$ and $|\!\downarrow\!(\bm{r},t)\rangle$ form local spin bases with quantization axis being $\bm{n}(\bm{r},t)$, the isospin $\bm{\mathcal{C}}$ can be pictured as a vector in the local frame moving with $\bm{n}(\bm{r}, t)$. In Appendix B, we have shown that in the particular gauge marked by $\chi=0$ (see Eq.~\eqref{eq:etavssApp}),
\begin{align}
 \bm{\mathcal{C}}=c_1\bm{\theta}+c_2\bm{\phi}+c_3\bm{n}=\frac1\xi(s_1\bm{\theta}+s_2\bm{\phi})+s_3\bm{n}, \label{eq:etavss}
\end{align}
where $\{\bm{\theta}, \bm{\phi}, \bm{n}\}$ form local bases associated with the local order parameter $\bm{n}(\bm{r},t)$ (Fig.~\ref{Fig:spinfootball}, left). Equation~\eqref{eq:etavss} indicates two important properties: (1) $\bm{\mathcal{C}}$ is coplanar with $\bm{n}$ and $\bm{s}$, which is specific to the particular gauge $\chi=0$; (2) while the isospin vector is constrained on the unit sphere $c_1^2+c_2^2+c_3^2=1$, the physical spin satisfies $\frac{s_1^2+s_2^2}{\xi^2}+s_3^2=1$, which constrains the tip of $\bm{s}$ on an prolate spheroid with semi-major axis being $\bm{n}(\bm{r},t)$ and semi-minor axis on its equator having length $\xi$ (Fig.~\ref{Fig:spinfootball}, right). The latter property is gauge independent and can be justified directly from Eq.~\eqref{eq:ds} without using Eq.~\eqref{eq:etavss} (see Appendix D), which, when written in gauge invariant from, becomes
\begin{equation}
  (\bm{s}\cdot\bm{n})^2+\frac{(\bm{s}\times\bm{n})^2}{\xi^2}=s_3^2+\frac{s_1^2+s_2^2}{\xi^2}=1. \label{eq:ellipsoid}
\end{equation}
For arbitrary gauges with $\chi\neq0$, it is easy to show that we always have $s_3=c_3$ and $s_1^2+s_2^2=\xi^2(c_1^2+c_2^2)$, but the angles between $s_{1,2}$ and $c_{1,2}$ will be different, \emph{i.e.}, $\bm{\mathcal{C}}$ will not be coplanar with $\bm{n}$ and $\bm{s}$.

Now, the physical picture of adiabatic spin evolution is clear: as the background order parameter $\bm{n}(\bm{r},t)$ moves slowly in space-time, the prolate spheroid moves with it. The motion of physical spin $\bm{s}$ is a superposition of the relative motion on the spheroid and the motion of the spheroid itself. The overall motion of $\bm{s}$ described by Eq.~\eqref{eq:ds} is purely geometrical as $\mathrm{d}t$ can be eliminated on both sides, as a result, a given path of $\bm{n}$ uniquely determines a path of $\bm{s}$ on the spheroid which is independent of the Hamiltonian. Associated with this geometric motion, a SU(2) Berry phase is accumulated by the electron wave function, which can be regarded as the (non-Abelian) gauge flux of a 't~Hooft-Polyakov monopole at the center of the unit sphere spanned by $\bm{n}$ (Table~\ref{Tab:comparison}).

A further remark: it seems to be a surprise that the magnitude of $\bm{s}$ varies on the spheroid since $\xi\le1$, but how can the physical spin have a nonconstant magnitude? We answer this question by studying the \emph{reduced} density matrix for the spin degree of freedom. It is a $2\times2$ matrix and can be written as $\rho_s=\frac12(1+\bm{a}\cdot\bm{\sigma})$, thus the expectation value of physical spin is $\bm{s}=\mathrm{Tr}[\rho_s\bm{\sigma}]=\bm{a}$. Now since $s^2\le1$, thus $a^2\le1$, and what follows is $\mathrm{Tr}\rho_s^2\le\mathrm{Tr}\rho_s$,  which suggests that the electron is \emph{effectively} in a mixed spin state. This can be attributed to the entanglement of spin and sublattice degrees of freedom, specifically, because $s_3=c_3$, we are able to infer the spin projection along $\bm{n}$ by measuring the probability difference on neighboring $A-B$ sites (vice versa). The entanglement provides us with partial information of spin orientation from the knowledge of sublattice, this destroys full coherence of the spin states.

\emph{Orbital dynamics}. In correspondence with the novel spin dynamics, the orbital dynamics of an individual electron also becomes non-trivial. By substituting Eq.~\eqref{eq:ds} into~\eqref{eq:dk} we get (see also Appendix B),
\begin{align}
  \dot{\bm{k}}&=(1-\xi^2)(\bm{s}\cdot\bm{n})(\bm{E}+\dot{\bm{r}}\times\bm{B}), \label{eq:SMF} \\
        \bm{E}&=\frac12\sin\theta(\partial_t\theta\nabla\phi-\nabla\theta\partial_t\phi), \label{eq:E} \\ \bm{B}&=-\frac12\sin\theta(\nabla\theta\times\nabla\phi), \label{eq:B}
\end{align}
the $\bm{E}$ and $\bm{B}$ fields here are the same as their FM counterparts, where they are responsible for the spin motive force~\cite{ref:SMF} and the topological Hall effect,~\cite{ref:THE} respectively. Also, as in FM systems, it is easy to check that Eqs.~\eqref{eq:E} and \eqref{eq:B} satisfy the Faraday's relation $\nabla\times\bm{E}+\frac{\partial\bm{B}}{\partial t}=0$.

However, quite different from the FM case, the gauge charge $\bm{s}\cdot\bm{n}$ in Eq.~\eqref{eq:SMF} is not just a constant, but involves the internal dynamics. In other words, the orbital motion is accompanied by a time-dependent gauge charge which should be determined by solving  the coupled equations Eqs.~\eqref{eq:EOMtotal} all together. Moreover, the factor $\xi^2$ results from the non-commutative term $2\bm{\mathcal{A}}^r_\mu\times\bm{\mathcal{A}}^r_\nu$ in Eq.~\eqref{eq:rrcurv}, it also reflects the coupling between spin and orbital dynamics. The parameter $\xi\in(0,1)$ plays a key role here: in the $\xi\rightarrow1$ limit, $1-\xi^2$ vanishes thus from Eqs.~\eqref{eq:ds} and~\eqref{eq:SMF} we get null results $\dot{\bm{s}}=0$ and $\dot{\bm{k}}=0$. In the other limit where $\xi\rightarrow0$, the solution of Eq.~\eqref{eq:ds} reduces to $\bm{s}=\pm\bm{n}$ if initial condition is $\bm{s}(0)=\pm\bm{n}(0)$, and Eq.~\eqref{eq:SMF} reduces to the FM Lorentz force equation, by which the system loses the manifest non-Abelian feature and behaves as two decoupled FM sub-systems. It deserves attention that in real AFM materials, both $A$ and $B$ sub-bands host majority carriers, but they are subject to effective Lorentz forces of opposite directions, which may lead to non-trivial spin transport.

Furthermore, the real space dynamics governed by Eq.~\eqref{eq:dr} exhibits spin-orbit coupling through the anomalous velocity term $\frac12(\bm{s}\times\bm{n})\cdot\dot{\bm{n}}\ \partial_{\bm{k}}\ln\xi$. It is along the same direction as $\partial_{\bm{k}}\varepsilon$, so Eq.~\eqref{eq:dr} amounts to give a modified group velocity. We mention that this term is unique to AFM  textures and has nothing to do with the anomalous velocity studied in FM or quantum Hall systems~\cite{ref:Adiabaticity}. It originates from the $\bm{\Omega}^{kr}_{\mu\nu}$ curvature that joints real space with BZ, the importance of which has been overlooked before. For better comparison, we summarize the fundamental electron dynamics of FM and AFM textures in Table~\ref{Tab:comparison}.

\begin{table}[]
 \begin{tabular}{l|l}
  \hline\hline
  FM spin texture & AFM spin texture \B\\
  \hline
  $\bm{s}=\bm{n}$                                    & $\ \dot{\bm{s}}=(1-\xi^2)(\bm{s}\cdot\bm{n})\dot{\bm{n}}$\\
  U(1) Abelian Berry Phase: & SU(2) non-Abelian Berry Phase:\\
  $\gamma(\Gamma)=\oint_{\Gamma}\mathcal{A}_\mu\mathrm{d}r_\mu$ & $U(\Gamma)=\mathcal{P}\exp[-i\oint_{\Gamma}\bm{\mathcal{A}}^r_\mu\cdot\bm{\tau}\mathrm{d}r_\mu]$\\
  Dirac monopole & 't Hooft-Polyakov monopole\\
  \hline
  $\dot{\bm{k}}=\bm{E}+\dot{\bm{r}}\times\bm{B} \qquad\qquad\ \ \ $     & $\ \dot{\bm{k}}=(1-\xi^2)(\bm{s}\cdot\bm{n})(\bm{E}+\dot{\bm{r}}\times\bm{B})$\\
  $\dot{\bm{r}}=-\partial_{\bm{k}}\varepsilon$ \B     & $\ \dot{\bm{r}}=-\partial_{\bm{k}}\varepsilon -\frac12(\bm{s}\times\bm{n})\cdot\dot{\bm{n}}\ \partial_{\bm{k}}\ln\xi$\\
  \hline\hline
 \end{tabular}
 \caption{Comparison of effective electron dynamics in FM and AFM textures. In the FM case, spin dynamics is trivial; along closed path $\Gamma$ the electron acquires an U(1) Berry phase which is the magnetic flux of a Dirac monopole; A Lorentz force is resulted in the orbital motion. In the AFM case, spin dynamics is non-trivial due to the mixture of degenerate sub-bands through a SU(2) non-Abelian Berry phase, which is the gauge flux generated by a 't Hooft-Polyakov monopole; The orbital dynamics is subject to a spin-dependent Lorentz force and an anomalous velocity.}{\label{Tab:comparison}}
\end{table}

A final remark on the theory part: In real materials with impurities, our fundamental equations~\eqref{eq:EOMtotal} are valid so long as spin coherence length is as large as, if not more than, the typical width of the texture. While this is quite true in FM materials, its validity in AFM materials awaits experimental verification. At extremely low temperatures, spin-flip scattering is dominated by magnetic impurities which can be made negligibly small in clean samples. Besides, spin-independent scattering processes (\emph{e.g.}, electron-phonon scattering) do not destroy our essential conclusions if $\dot{\bm{r}}$ is understood as the drift velocity of carriers. We mention that AFM spintronics is an emerging field where very little is known. While it shares some similarities with the established FM spintronics, it is not always correct to copy ideas from FM systems. The adiabatic electron dynamics studied in this paper is one example.

\section{IV Examples}

\begin{figure}[]
   \centering
   \includegraphics[width=0.88\linewidth]{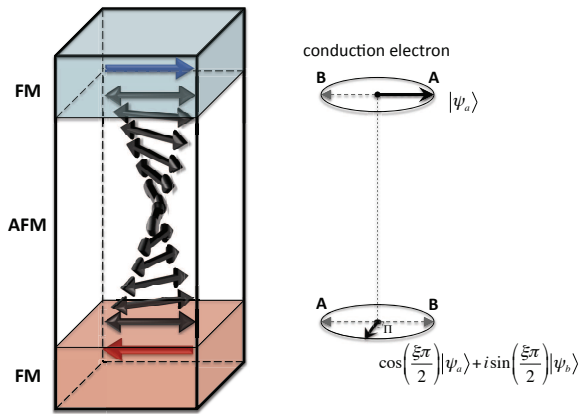}
   \caption{(Color online) Left: FM/AFM/FM trilayer with opposite FM orientations on two sides. The black double arrows represent the $A$-$B$ sublattices of the AFM layer, which is dragged into a spiraling texture due to exchange bias on the interfaces. Right: incoming electrons only enter the $A$ sub-band due to the upper FM polarizer, the out-going electrons partially occupy the $B$ sub-band depending on the value of $\xi$.}{\label{Fig:DW}}
\end{figure}

First, consider a spiraling AFM texture sandwiched by two ferromagnetic layers (see Fig.~\ref{Fig:DW}). This magnetic structure has been realized in Co/FeMn/Py trilayers in a recent experiment~\cite{ref:SpiralingSpin}, where the FM order of Co layer is nearly fixed but that of Py can be rotated by external magnetic field. The AFM order is dragged into a spiral due to the exchange bias effect on the AFM/FM interfaces. The layer thickness of FeMn is roughly $10\sim20\ \mathrm{nm}$ and can be made even larger, which far exceeds the lattice constant thus adiabatic approximation is valid; Meanwhile, typical spin coherence length is larger than the layer thickness at low temperatures so that spin evolution is governed by Eq.~\eqref{eq:ds}.

When an electron flows from top to bottom with applied current, the top FM layer polarizes its spin so that it enters the $A$ sub-band across the interface. According to Eq.~\eqref{eq:ds}, the physical spin orientation of the electron after passing through the AFM layer is rotated by $\Pi=\pi-\arctan[\xi\tan\xi\pi]$ if $\xi<\frac12$, and $\Pi=-\arctan[\xi\tan\xi\pi]$ is $\xi>\frac12$. This is a topological result that only depends on the initial and final directions of $\bm{n}$, but is \textit{independent} of the texture's profile detail. When $\xi\rightarrow0$, $\Pi$ reduces to $\pi$, which means the electron spin follows $\bm{n}$ and remains in the $A$ sub-band, thus it flows into the bottom FM layer with a lower resistance; in the $\xi\rightarrow 1$ limit, $\Pi$ vanishes and the electron completely evolves into the $B$ sub-band thus experiencing a higher resistance. For an arbitrary $\xi$ and an arbitrary total rotation of the spiral denoted by $\Phi$, the electron will partially evolve into the $B$ sub-band with the wave function $\cos(\xi\Phi/2)|\psi_a\rangle+ i\sin(\xi\Phi/2)|\psi_b\rangle$, thus the total resistance is
\begin{equation}
 \rho=\rho_0+\frac12\Delta\rho[1-\cos(\xi\Phi)],
 \label{eq:MR}
\end{equation}
where $\rho_0$ is the intrinsic resistance of the AFM texture itself, which depends monotonously but not too much on $\Phi$. $\Delta\rho$ represents the magnetoresistance of the spin valve which is determined by material details of the two FM layers and is independent of $\Phi$. If $\Phi$ is increased beyond $\pi$, $\rho$ will reach a maximum at $\Phi_m=\pi/\xi$ and then reduces. The resistance maximum, if observed, serves as an experimental verification of Eq.~\eqref{eq:ds}. Moreover, measuring $\Phi_m$ also enables us to find $\xi$ without calculating the band structure.

We remark that the above results survive in the presence of diffusive processes so long as spin-flip scattering is ignored. The reason is that spin-independent scattering only deflects $\bm{k}$-space orbit, whereas the $\bm{s}$ dynamics is determined by the variation of $\bm{n}$ that is blind to $\bm{k}$ in one dimension. In addition, FeMn is a non-collinear antiferromagnet that has more than two sub-lattices. To test our theory unambiguously, we can replace FeMn by the collinear IrMn which is feasible for current technique~\cite{ref:private}. Moreover, we are aware of the experiment~\cite{ref:Mnlayer} where the spiraling AFM texture exhibits spatial \textit{periodic} patterns, it provides a better way of realizing large $\Phi's$.

\begin{figure}[]
   \centering
   \includegraphics[height=0.23\textheight]{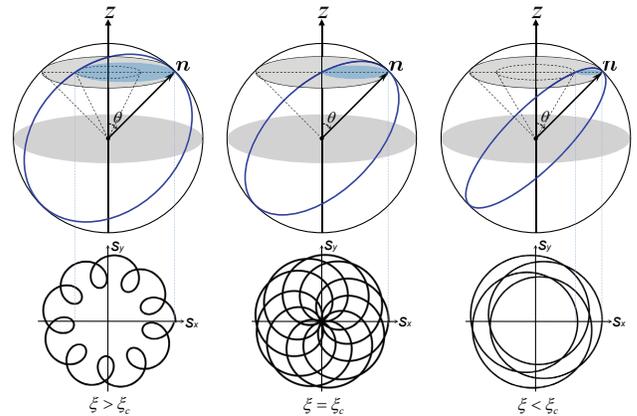}
   \caption{(Color online) Spin evolutions for three different $\xi$'s when $\bm{n}(t)$ is moving round a cone with constant angle $\theta$ from the $z$ axis. Upper panels: the tip of $\bm{s}$ respects two constraints: it stays both on the cone's bottom (small gray slab) and on the spheroid described by Eq.~\eqref{eq:ellipsoid} (blue ellipsoid), thus the vector $\bm{s}$ is confined in between two cones with different semiangles. Lower panels: orbits of the tip from bird's eye view. The topology of the orbits is separated into two classes (left and right) by the critical case (middle) where the inner cone's semiangle shrinks to zero. The orbits are not necessarily commensurate with $\bm{n}$.} {\label{Fig:Rings}}
\end{figure}

Consider a second example where $\bm{n}(t)$ is varying round a cone of constant semiangle $\theta$ in the laboratory frame, which can be realized in a spin wave (see Fig.~\ref{Fig:Rings}). According to Eq.~\eqref{eq:ds}, we know that $\mathrm{d}s_z=0$ due to $\mathrm{d}n_z=0$, thus, the tip of $\bm{s}$ should stay in the bottom plane of the cone. On the other hand, we learn from Eq.~\eqref{eq:ellipsoid} that the tip is constrained on the spheroid that moves with the instantaneous $\bm{n}(t)$. Therefore, the actual orbit traversed by the tip is contained in the intersection of the two constraints. Through some straightforward geometric analysis, we know that $\bm{s}$ is bounded between the $\bm{n}$ cone and an inner cone whose semiangle depends on $\xi$. Figure~\ref{Fig:Rings} depicts the actual orbits of $\bm{s}$ for three different $\xi$'s: they all exhibit precession and nutation, which can be easily read out from the bird's eye view. Remarkably, the motion of $\bm{s}$ falls into two topologically distinct classes separated by the critical condition
\begin{align}
 \xi_c^2=\frac{\cos^2\theta}{(1+\cos^2\theta)}.
\end{align}
In a real spin wave, $\theta$ is nearly zero, thus $\xi_c\approx1/\sqrt{2}$. For real materials, we expect $t\le J$, thus from Eq.~\eqref{eq:overlap} we know that for a partially filled band, $\xi$ is always smaller than $1/\sqrt{2}$. Therefore, the $\xi<\xi_c$ phase is more realistic.

\section{Conclusions}

In this paper, we find that a slowly varying AFM texture renders adiabatic dynamics of conduction electrons, which are described by three coupled equations of motion [Eqs.~\eqref{eq:EOMtotal}]. Quite different from the FM case, the adiabaticity in AFM materials does not imply strict alignment between conduction electron spins and the profile of background texture. Instead, the adiabatic spin evolution is a superposition of a motion following the background order plus a motion on a prolate spheroid attached to the local order, where the latter originates from internal dynamics between degenerate bands. The overall motion of the spin is still geometric; it can be attributed to the accumulation of a SU(2) non-Abelian Berry phase originating from the gauge flux of an effective 't~Hooft-Polyakov monopole in the parameter space.

The corresponding orbital dynamics shares some similarities with FM materials in that the $\bm{k}$-space dynamics can be described by an effective Lorentz force equation. However, two prominent differences in the orbital dynamics distinguish an AFM system from its FM counterpart: first, the gauge charge is dynamical rather than constant, by which spin and orbital motions no longer separate; second, the group velocity is renormalized by a spin-dependent anomalous velocity, which is quite different from what has been studied before.

Our theory lays the foundation for charge and spin transports in AFM texture systems, which will be applied to real materials in the future. The validity of our theory needs to be tested experimentally since available data on AFM spintronic materials are very rare. Theoretically, this paper solves only the first half of the whole story; the other half, \emph{i.e.}, the converse effect regarding the back-reaction of current on background AFM order, will appear in a forthcoming paper.

\section{Acknowledgements}
We are grateful for Dr. Yizhuang You and Prof. Biao Wu for numerous helps on detailed calculations, and Dr. Karin Everschor for insightful comments. We thank Prof. Maxim Tsoi and Prof. Jianwang Cai for discussions on possible experimental observations. We also acknowledge helps from J.~Zhou, S.~A.~Yang, A.~H.~MacDonald, X.~Li, Y.~Gao, and D.~Xiao for comments from diverse aspects. This work is supported by DOE-DMSE, NBRPC, NSFC, and the Welch Foundation.

\appendix
\section{}
For simplicity, we denote the joint space-time coordinate as $r_\mu\equiv(t,\bm{r}_c)$, so that the local Hamiltonian $\mathcal{H}(r_\mu)$ is parameterized by $r_\mu$ and the wave packet of the electron can be expressed as
\begin{align}
 |W&(r_\mu,\bm{k}_c)\rangle\!=\!\!\int\mathrm{d}\bm{k}w(\bm{k}-\bm{k}_c)[c_a|\psi_a\rangle +c_b|\psi_b\rangle] \notag\\
 &=\!\int\!\mathrm{d}\bm{k}w(\bm{k}-\bm{k}_c)[c_a|A(\bm{k})\rangle|\!\uparrow\!(r_\mu)\rangle+c_b|B(\bm{k})\rangle|\!\downarrow\!(r_\mu)\rangle], \label{eq:wavepacket}
\end{align}
where the profile function $w(\bm{k}-\bm{k}_c)$ satisfies $\int\mathrm{d}\bm{k}|w(\bm{k}-\bm{k}_c)|^2=1$ and $\int\mathrm{d}\bm{k}\bm{k}|w(\bm{k}-\bm{k}_c)|^2=\bm{k}_c$. The coefficients $c_a$ and $c_b$ reflect relative contributions from the two sub-bands. Here, we define the isospinor $\tilde{c}=[c_a,c_b]^T$, then the normalization condition $|c_a|^2+|c_b|^2=1$ becomes $\tilde{c}^\dagger\tilde{c}=1$.~\cite{ref:noteisospin} The effective Lagrangian for the wave packet is then written as~\cite{ref:Dimi}
\begin{align}
  \mathcal{L}&=\langle W|(i\frac{\partial}{\partial t}-\mathcal{H})|W\rangle=\mathcal{L}(r_\mu, k_\mu, \tilde{c};\ \dot{r}_\mu, \dot{k}_\mu, \dot{\tilde{c}}) \notag\\
  &=\varepsilon+\bm{k}_c\!\cdot\dot{\bm{r}}_c+i\tilde{c}^{\dagger}\dot{\tilde{c}} +\tilde{c}^\dagger(A^r_\mu\dot{r}_\mu+A^k_\mu\dot{k}_\mu)\tilde{c}, \label{eq:Lagrangian}
\end{align}
where $k_\mu=(0,\bm{k}_c)$ has no temporal component in contrast to $r_\mu=(t,\bm{r}_c)$, but it is still written this way just to simplify symbols. Here the Berry connections are $2\times2$ matrices and are functions of both $r_\mu$ and $k_\mu$. The real space components are defined by~\cite{ref:Dimi,ref:Dalibard,ref:NABerryPhase}
\begin{align}
 A^r_\mu&=i \begin{bmatrix}
            \langle u_a|\partial^r_\mu|u_a\rangle, & \langle u_a|\partial^r_\mu|u_b\rangle\\
            \langle u_b|\partial^r_\mu|u_a\rangle, & \langle u_b|\partial^r_\mu|u_b\rangle
            \end{bmatrix} \notag\\
        &=\!\frac12\!\begin{bmatrix}
            \cos\theta\partial_\mu\phi & \xi(-i\partial_\mu\theta-\sin\theta\partial_\mu\phi) \\
            \xi(i\partial_\mu\theta-\sin\theta\partial_\mu\phi) & -\cos\theta\partial_\mu\phi
            \end{bmatrix} \notag\\
        &=\frac12[-\tau_1\xi\sin\theta\partial_\mu\phi +\tau_2\xi\partial_\mu\theta +\tau_3\cos\theta\partial_\mu\phi],
\end{align}
where $\{\tau_1, \tau_2, \tau_3\}\equiv\bm{\tau}$ is a vector of Pauli matrices representing isospin, it should not be confused with the real spin operator $\bm{\sigma}$. Similarly, the momentum space components are
\begin{align}
 A^k_\mu&=i \begin{bmatrix}
            \langle u_a|\partial^k_\mu|u_a\rangle, & \langle u_a|\partial^k_\mu|u_b\rangle\\
            \langle u_b|\partial^k_\mu|u_a\rangle, & \langle u_b|\partial^k_\mu|u_b\rangle
            \end{bmatrix}
         =  \begin{bmatrix}
            0 & 0\\
            0 & 0
            \end{bmatrix},
\end{align}
which in general should not vanish if spin-orbit coupling terms are added to the original Hamiltonian. To avoid complicated manipulations on matrix products in the final results, we define the isospin vector
\begin{align}
 \bm{\mathcal{C}}&=\{c_1, c_2, c_3\}=\tilde{c}^{\dagger}\bm{\tau}\tilde{c} \notag\\
 &=\{2\mathrm{Re}(c_a c_b^*), -2\mathrm{Im}(c_a c_b^*), |c_a|^2-|c_b|^2\}, \label{eq:eta}
\end{align}
with which the Berry connection $A^r_\mu$ can be expressed as a vector in the isospin vector space (the adjoint representation),
\begin{align}
\bm{\mathcal{A}}^r_\mu&=\frac12\mathrm{Tr}[\bm{\tau}A^r_\mu]=\{(\mathcal{A}^r_\mu)^1,\ (\mathcal{A}^r_\mu)^2,\ (\mathcal{A}^r_\mu)^3\} \notag\\
&=\frac12\{-\xi\sin\theta\partial_\mu\phi,\ \xi\partial_\mu\theta,\ \cos\theta\partial_\mu\phi\}, \label{eq:Amudecomposition}
\end{align}
where we have used $\frac12\mathrm{Tr}[\tau_i\tau_j]=\delta_{ij}$. In a similar way, $\bm{\mathcal{A}}^k_\mu=\frac12\mathrm{Tr}[\bm{\tau}A^k_\mu]=0$, but we still keep it at this step for the completeness of the entire formalism. The effective Lagrangian now becomes
\begin{align}
 \mathcal{L}=\varepsilon+k_\mu \dot{r}_\mu+i\tilde{c}^{\dagger}\dot{\tilde{c}}+ \bm{\mathcal{C}}\cdot[\bm{\mathcal{A}}^r_\mu\dot{r}_\mu+\bm{\mathcal{A}}^k_\mu\dot{k}_\mu], \label{eq:Ladjoint}
\end{align}
where the $i\tilde{c}^{\dagger}\dot{\tilde{c}}$ term cannot be expressed in terms of $\bm{\mathcal{C}}$ and $\dot{\bm{\mathcal{C}}}$, but this poses no problem in the following. Both Eqs.~\eqref{eq:Lagrangian} and~\eqref{eq:Ladjoint} are useful.

To obtain the equations of motion for the three parameters $(\bm{\mathcal{C}}, r_\mu, k_\mu)$, we resort to the following variational principles.

\emph{(1)} $\delta\mathcal{L}/\delta \tilde{c}=0$ on Eq.~\eqref{eq:Lagrangian} gives:
\begin{subequations}
 \begin{align}
  \dot{\tilde{c}}&=i[A^r_\mu\dot{r}_\mu+A^k_\mu\dot{k}_\mu]\tilde{c} \\
  \dot{\tilde{c}}^\dagger&=-i\tilde{c}^\dagger[A^r_\mu\dot{r}_\mu+A^k_\mu\dot{k}_\mu]
 \end{align}
\end{subequations}
then from $\bm{\mathcal{C}}=\tilde{c}^{\dagger}\bm{\tau}\tilde{c}$ we have,
\begin{align}
 \dot{\bm{\mathcal{C}}}&=\dot{\tilde{c}}^\dagger\bm{\tau}\tilde{c}+\tilde{c}^\dagger\bm{\tau}\dot{\tilde{c}} \notag\\
 &=i\dot{r}_\mu\tilde{c}^\dagger[\bm{\tau}A^r_\mu-A^r_\mu\bm{\tau}]\tilde{c} +i\dot{k}_\mu\tilde{c}^\dagger[\bm{\tau}A^k_\mu-A^k_\mu\bm{\tau}]\tilde{c} \label{eq:etavari}
\end{align}
where $\bm{\tau}A_\mu$ terms are matrix products. In view of the decomposition Eq.~\eqref{eq:Amudecomposition}, we take a specified component of Eq.~\eqref{eq:etavari},
\begin{align}
 \dot{\mathcal{C}}_\alpha&=i\dot{r}_\mu(\mathcal{A}^r_\mu)^\beta\tilde{c}^\dagger(\tau_\alpha\tau_\beta-\tau_\beta\tau_\alpha)\tilde{c} \notag\\ &\qquad+i\dot{k}_\mu(\mathcal{A}^k_\mu)^\beta\tilde{c}^\dagger(\tau_\alpha\tau_\beta-\tau_\beta\tau_\alpha)\tilde{c} \notag\\
 &=-2\varepsilon_{\alpha\beta\gamma}[(\mathcal{A}^r_\mu)^\beta\dot{r}_\mu+(\mathcal{A}^k_\mu)^\beta\dot{k}_\mu](\tilde{c}^\dagger\tau_\gamma\tilde{c}),
\end{align}
when written in the isospin vector form, the above equation becomes
\begin{align}
 \dot{\bm{\mathcal{C}}}=2\bm{\mathcal{C}}\times(\bm{\mathcal{A}}^r_\mu\dot{r}_\mu +\bm{\mathcal{A}}^k_\mu\dot{k}_\mu) \label{eq:EOMetaApp}
\end{align}
which proves Eq.~\eqref{eq:EOMeta}.

\emph{(2)} $\delta\mathcal{L}/\delta r_\mu=0$ on Eq.~\eqref{eq:Ladjoint} requires some care:
\begin{subequations}
\begin{align}
 \frac{\delta\mathcal{L}}{\delta r_\mu}&=\frac{\partial\mathcal{L}}{\partial r_\mu} -\frac{\mathrm{d}}{\mathrm{d}t}\frac{\partial\mathcal{L}}{\partial \dot{r}_\mu}=0\ \ \mbox{with,} \notag \\
 \frac{\partial\mathcal{L}}{\partial r_\mu}&=\partial^r_\mu\varepsilon+\bm{\mathcal{C}}\cdot[(\partial^r_\mu\bm{\mathcal{A}}^r_\nu)\dot{r}_\nu +(\partial^r_\mu\bm{\mathcal{A}}^k_\nu)\dot{k}_\nu] \\
 \frac{\mathrm{d}}{\mathrm{d}t}\frac{\partial\mathcal{L}}{\partial \dot{r}_\mu}\!&=\dot{k}_\mu+(\dot{\bm{\mathcal{C}}}\cdot\bm{\mathcal{A}}^r_\mu +\bm{\mathcal{C}}\cdot\frac{\mathrm{d}}{\mathrm{d}t}\bm{\mathcal{A}}^r_\mu) \notag\\
 &=\dot{k}_\mu+2[\dot{r}_\nu\bm{\mathcal{C}}\cdot(\bm{\mathcal{A}}^r_\nu\times\bm{\mathcal{A}}^r_\mu) +\dot{k}_\nu\bm{\mathcal{C}}\cdot(\bm{\mathcal{A}}^k_\nu\times\bm{\mathcal{A}}^r_\mu)] \notag\\
 &\qquad\ \ +\bm{\mathcal{C}}\cdot[(\partial^r_\nu\bm{\mathcal{A}}^r_\mu)\dot{r}_\nu +(\partial^k_\nu\bm{\mathcal{A}}^r_\mu)\dot{k}_\nu]
\end{align}
\end{subequations}
where in the last line Eq.~\eqref{eq:EOMetaApp} has been used. Hence we obtain the equation of motion
\begin{align}
 \dot{k}_\mu=\partial^r_\mu\varepsilon +\bm{\mathcal{C}}\cdot[\bm{\Omega}^{rr}_{\mu\nu}\dot{r}_\nu+\bm{\Omega}^{rk}_{\mu\nu}\dot{k}_\nu], \label{eq:EOMkApp}
\end{align}
where the Berry curvatures are defined as
\begin{subequations}
\label{eq:BerrycurvatureRR}
\begin{align}
  \bm{\Omega}^{rr}_{\mu\nu}&\equiv\partial^r_\mu \bm{\mathcal{A}}^r_\nu-\partial^r_\nu\bm{\mathcal{A}}^r_\mu +2\bm{\mathcal{A}}^r_\mu\times\bm{\mathcal{A}}^r_\nu, \\
  \bm{\Omega}^{rk}_{\mu\nu}&\equiv\partial^r_\mu \bm{\mathcal{A}}^k_\nu-\partial^k_\nu\bm{\mathcal{A}}^r_\mu +2\bm{\mathcal{A}}^r_\mu\times\bm{\mathcal{A}}^k_\nu,
\end{align}
\end{subequations}
they are antisymmetric tensors with permutations of $r_\mu$ and $k_\mu$, and at the same time they are vectors in the isospin vector space -- the internal space unique to non-Abelian gauge theory.

\emph{(3)} $\delta\mathcal{L}/\delta k_\mu=0$ on Eq.~\eqref{eq:Ladjoint} follows quite similar procedures as above, and the equation of motion is:
\begin{align}
  \dot{r}_\mu& = -\partial^k_\mu\varepsilon -\bm{\mathcal{C}}\cdot[\bm{\Omega}^{kr}_{\mu\nu}\dot{r}_\nu+\bm{\Omega}^{kk}_{\mu\nu}\dot{k}_\nu], \label{eq:EOMrApp}
\end{align}
where the Berry curvatures are
\begin{subequations}
\label{eq:BerrycurvatureKK}
\begin{align}
  \bm{\Omega}^{kk}_{\mu\nu}&\equiv\partial^k_\mu \bm{\mathcal{A}}^k_\nu-\partial^k_\nu\bm{\mathcal{A}}^k_\mu +2\bm{\mathcal{A}}^r_\mu\times\bm{\mathcal{A}}^r_\nu, \\
  \bm{\Omega}^{kr}_{\mu\nu}&\equiv\partial^k_\mu \bm{\mathcal{A}}^r_\nu-\partial^r_\nu\bm{\mathcal{A}}^k_\mu +2\bm{\mathcal{A}}^k_\mu\times\bm{\mathcal{A}}^r_\nu,
\end{align}
\end{subequations}
which, together with Eqs.~\eqref{eq:BerrycurvatureRR}, form a generalized matrix of Berry curvature jointing real space and BZ into a unified parameter space,
\begin{align}
 \tilde{\bm{\Omega}}_{\mu\nu}=
 \begin{bmatrix}
 \bm{\Omega}^{rr}_{\mu\nu} & \bm{\Omega}^{rk}_{\mu\nu}\\
 \bm{\Omega}^{kr}_{\mu\nu} & \bm{\Omega}^{kk}_{\mu\nu}
 \end{bmatrix}.
\end{align}
Eqs.~\eqref{eq:EOMkApp},~\eqref{eq:EOMrApp}, and Eq.~\eqref{eq:EOMetaApp} justify the main conclusions in Sec.~II. It is worth mentioning that the Abelian version of the above formalism is a well-established field, whose great success has been proved by compelling evidences in the last decade.~\cite{ref:Adiabaticity} However, the non-Abelian version is still underdeveloped in previous work.~\cite{ref:Dimi}

\section{}

Before deriving Eqs.~\eqref{eq:EOMtotal}, special attention should be paid on the fact that gauge fields (Berry curvatures) in non-Abelian theory are not gauge invariant, but gauge covariant. It is the isospin \emph{scalars} $\bm{\mathcal{C}}\cdot\bm{\Omega}_{\mu\nu}$ appearing in Eqs.~\eqref{eq:EOM} that respect gauge invariance. Specifically, as we make a gauge transformation on the wave functions $|\psi_a\rangle$ and $|\psi_b\rangle$, change of $\bm{\Omega}$ just compensates that of $\bm{\mathcal{C}}$. However, to perform real calculations, we have to work in a specified gauge. In this paper, the gauge is fixed in the choice of local spin eigenstates, which are obtained by acting $U(\bm{r}, t)=e^{-i\sigma_z\phi/2}e^{-i\sigma_y\theta/2}e^{-i\sigma_z\chi/2}$ on the eigenstates of $\sigma_z$. While $\theta(\bm{r},t)$ and $\phi(\bm{r},t)$ are physical, $\chi(\bm{r}, t)$ is not and can be chosen arbitrarily; the gauge is fixed by setting $\chi=0$.

From Eq.~\eqref{eq:Amudecomposition} the real-space curvature is obtained:
\begin{align}
  \bm{\Omega}^{rr}_{\mu\nu}&\equiv\partial^r_\mu \bm{\mathcal{A}}^r_\nu-\partial^r_\nu\bm{\mathcal{A}}^r_\mu +2\bm{\mathcal{A}}^r_\mu\times\bm{\mathcal{A}}^r_\nu \notag\\
  &=\{0, 0, (\xi^2-1)\frac12\sin\theta(\partial^r_\mu\theta\partial^r_\nu\phi-\partial^r_\nu\theta\partial^r_\mu\phi)\} \notag\\
  &=\{0,\ 0,\ (\xi^2-1)\frac12\bm{n}\cdot(\partial^r_\mu\bm{n}\times\partial^r_\nu\bm{n})\}, \label{eq:rrcurvature}
\end{align}
where $\bm{n}=\{\sin\theta\cos\phi, \sin\theta\sin\phi, \cos\theta\}$ is the local order parameter. We see that only the third component is non-zero in our particular gauge marked by $\chi=0$. But one can check that in any gauge with $\chi\neq0$, the first two components do not vanish. However, the third component is actually gauge invariant and it has the form of skyrmion density. The cross components of Berry curvature are obtained in a similar way,
\begin{align}
  \bm{\Omega}^{rk}_{\mu\nu}=-\bm{\Omega}^{kr}_{\nu\mu}&=\partial^r_\mu \bm{\mathcal{A}}^k_\nu-\partial^k_\nu\bm{\mathcal{A}}^r_\mu +2\bm{\mathcal{A}}^r_\mu\times\bm{\mathcal{A}}^k_\nu \notag\\
  &=\frac12\{\partial^k_\nu\xi\sin\theta\partial^r_\mu\phi\ -\partial^k_\nu\xi\partial^r_\mu\theta,\ 0\}, \label{eq:rkcurvature}
\end{align}
where again the first two components are changeable subject to gauge transformations, whereas the third is gauge invariant. Moreover, due to $\bm{\mathcal{A}}^k_\mu=0$, the BZ space Berry curvature $\bm{\Omega}^{kk}_{\mu\nu}$ vanishes and we will not mention it
 in the following. Now let us turn to the equations of motion.

\emph{(1)} Substitute Eq.~\eqref{eq:Amudecomposition} into Eq.~\eqref{eq:EOMetaApp}, noting that $\dot{\theta}=\dot{r}_\mu\partial_\mu\theta=\partial_t\theta+\dot{\bm{r}}_c\!\cdot\!\nabla\theta$ and the same for $\dot{\phi}$,
\begin{align}
 \frac{\mathrm{d}}{\mathrm{d}t}
 \begin{bmatrix}
  c_1\\ c_2\\ c_3
 \end{bmatrix}
=\begin{bmatrix}
 0 & \cos\theta\dot{\phi} & -\xi\dot{\theta} \\
 -\cos\theta\dot{\phi} & 0 & -\xi\sin\theta\dot{\phi} \\
 \xi\dot{\theta} & \xi\sin\theta\dot{\phi} & 0
 \end{bmatrix}
 \begin{bmatrix}
  c_1\\ c_2\\ c_3
 \end{bmatrix}. \label{eq:etadyn}
\end{align}
Our target is to transform the dynamics of $\bm{\mathcal{C}}$ to the dynamics of the physical spin defined by
\begin{align}
\bm{s}=\langle W(r_\mu)|\bm{\sigma}|W(r_\mu)\rangle, \label{eq:physicalspin}
\end{align}
which respects gauge invariance. From Eqs.~\eqref{eq:wavefunctions}, \eqref{eq:wavepacket}, and \eqref{eq:eta}, and \eqref{eq:physicalspin} we know the components of $\bm{s}$ in the \emph{lab} frame after some tedious algebra,
\begin{subequations}
\label{eq:sn}
\begin{align}
  s_x&=c_3\sin\theta\cos\phi+\xi[c_1\cos\theta\cos\phi-c_2\sin\phi],\\
  s_y&=c_3\sin\theta\sin\phi+\xi[c_1\cos\theta\sin\phi+c_2\cos\phi],\\
  s_z&=c_3\cos\theta-\xi c_1\sin\theta,
\end{align}
\end{subequations}
then we take the total time derivative over each component of $\bm{s}$, for example,
\begin{align}
 \dot{s}_x=\ &\dot{c}_3\sin\theta\cos\phi+c_3(\cos\theta\cos\phi\dot{\theta}-\sin\theta\sin\phi\dot{\phi}) \notag\\
 &+\xi[\dot{c}_1\cos\theta\cos\phi-c_1(\sin\theta\cos\phi\dot{\theta}+\cos\theta\sin\phi\dot{\phi})] \notag\\
 &\qquad-\xi[\dot{c_2}\sin\phi+c_2\cos\phi\dot{\phi}] \notag\\
 =\ &c_3(1-\xi^2)(\cos\theta\cos\phi\dot{\theta}-\sin\theta\sin\phi\dot{\phi}) \notag\\
 =\ &c_3(1-\xi^2)\dot{n}_x \label{eq:sx}
\end{align}
where in deriving the second equality above Eq.~\eqref{eq:etadyn} has been used. Similarly,
\begin{align}
 \dot{s}_y=c_3(1-\xi^2)\dot{n}_y,\quad \dot{s}_z=c_3(1-\xi^2)\dot{n}_z. \label{eq:syz}
\end{align}
To eliminate $c_3$ in the above equations, we reverse Eqs.~\eqref{eq:sn} and obtain,
\begin{align}
 c_3=s_x\sin\theta\cos\phi+s_y\sin\theta\sin\phi+s_z\cos\theta=\bm{s}\cdot\bm{n}
\end{align}
then from Eqs.~\eqref{eq:sx} and~\eqref{eq:syz}, we obtain a simple and elegant equation of motion for the physical spin,
\begin{align}
     \dot{\bm{s}} = (1-\xi^2)(\bm{s}\cdot\bm{n})\dot{\bm{n}}, \label{eq:dsApp}
\end{align}
which justifies Eq.~\eqref{eq:ds}.

\emph{(2)}
To justify Eq.~\eqref{eq:dk}, we substitute Eqs.~\eqref{eq:rrcurvature} and ~\eqref{eq:rkcurvature} into Eq.~\eqref{eq:EOMkApp}, regarding that $k_\mu$ has only spatial but no temporal components, we arrive at,
\begin{align}
 \dot{\bm{k}}_c&=\frac12c_3(\xi^2-1)\bm{n}\cdot(\nabla\bm{n}\times\dot{\bm{n}}) +\frac12\dot{\xi}[c_1\sin\theta\nabla\phi-c_2\nabla\theta]\notag\\
 &=\frac12\bm{n}\cdot\{\nabla\bm{n}\times[(\xi^2-1)(\bm{s}\cdot\bm{n})\dot{\bm{n}}]\} \notag\\
 &=-\frac12\bm{n}\cdot(\nabla\bm{n}\times\dot{\bm{s}}) \label{eq:dkApp}
\end{align}
where $\dot{\xi}=0$ and Eq.~\eqref{eq:dsApp} have been used. We also have ignored $\partial^r_\mu\varepsilon$ term in Eq.~\eqref{eq:dkApp} since the band structure is only a function of $\bm{k}$ and independent of space-time in the adiabatic approximation. As in Sec.~III, we will omit the superscript $c$ in the following for simplicity. To make better comparisons with the spin motive force~\cite{ref:Shengyuan,ref:SMF} and the topological Hall effect~\cite{ref:THE} discovered in FM materials, we also derive another suggestive form of Eq.~\eqref{eq:dk}. Take an arbitrary component $i$ of Eq.~\eqref{eq:dkApp},
\begin{align}
\dot{k}_i&=\frac12c_3(\xi^2-1)\sin\theta(\partial_i\theta\dot{\phi}-\dot{\theta}\partial_i\phi) \notag\\
&=\frac12c_3(\xi^2-1)\sin\theta\{[\partial_i\theta\partial_t\phi-\partial_t\theta\partial_i\phi] \notag\\ &\qquad\qquad\qquad+[\partial_i\theta(\dot{r}_j\partial_j\phi)-(\dot{r}_j\partial_j\theta)\partial_i\phi]\} \notag\\
&=\frac12c_3(\xi^2-1)\{\sin\theta[\partial_i\theta\partial_t\phi-\partial_t\theta\partial_i\phi] \notag\\
&\qquad\qquad+\sin\theta\varepsilon_{ijk}\varepsilon_{klm}\dot{r}_j\partial_l\theta\partial_m\phi \} \label{eq:Lorentz}
\end{align}
where $\dot{\theta}=\partial_t\theta+\dot{r}_i\partial_i\theta$ (the same for $\dot{\phi}$) and the identity $\varepsilon_{ijk}\varepsilon_{klm}=\delta_{il}\delta_{jm}-\delta_{im}\delta_{jl}$ have been used. Eq.~\eqref{eq:Lorentz} can be written in a concise way as,
\begin{align}
  \dot{\bm{k}}=&(1-\xi^2)(\bm{s}\cdot\bm{n})(\bm{E}+\dot{\bm{r}}\times\bm{B}), \\
        \bm{E}&=\frac12\sin\theta(\partial_t\theta\nabla\phi-\nabla\theta\partial_t\phi), \notag \\
        \bm{B}&=-\frac12\sin\theta(\nabla\theta\times\nabla\phi),
\end{align}
which proves Eqs.~\eqref{eq:SMF} -- \eqref{eq:B} in Sec.~III.

\emph{(3)} From Eq.~\eqref{eq:EOMrApp} and Eq.~\eqref{eq:rkcurvature}, we have,
\begin{align}
 \dot{\bm{r}}=-\partial_{\bm{k}}\varepsilon+\frac12\partial_{\bm{k}}\xi(c_1\sin\theta\dot{\phi}-c_2\dot{\theta}). \label{eq:anomalousvelocity}
\end{align}
There is a smart way to eliminate $c_{1,2}$ in terms of physical spin. Notice that in the special gauge marked by $\chi=0$, the isospin vector $\bm{\mathcal{C}}$ can be pictured as a vector in the local frame extended by $\bm{\theta}$, $\bm{\phi}$, and $\bm{n}$ (See Fig.~\ref{Fig:spinfootball}): $\bm{\mathcal{C}}=c_1\bm{\theta}+c_2\bm{\phi}+c_3\bm{n}$, and components of the real spin $\bm{s}$ in this local frame are
\begin{align}
 s_1&=s_x\cos\theta\cos\phi+s_y\cos\theta\sin\phi-s_z\sin\theta=\xi c_1 \\
 s_2&=-s_x\sin\phi+s_y\cos\phi=\xi c_2 \\
 s_3&=s_x\sin\theta\cos\phi+s_y\sin\theta\sin\phi+s_z\cos\theta=c_3
\end{align}
where Eqs.~\eqref{eq:sn} have been used, and we obtain the important relation,
\begin{align}
 \bm{s}=\xi(c_1\bm{\theta}+c_2\bm{\phi})+c_3\bm{n}. \label{eq:etavssApp}
\end{align}
Since $\dot{\bm{n}}=\dot{\theta}\bm{\theta}+\sin\theta\dot{\phi}\bm{\phi}$, we immediately have
\begin{align}
\bm{n}\cdot(\bm{s}\times\dot{\bm{n}})=\xi(c_1\sin\theta\dot{\phi}-c_2\dot{\theta}),
\end{align}
thus Eq.~\eqref{eq:anomalousvelocity} becomes
\begin{align}
 \dot{\bm{r}}=-\partial_{\bm{k}}\varepsilon-\frac12(\bm{s}\times\bm{n})\cdot\dot{\bm{n}}\partial_{\bm{k}}\ln\xi
\end{align}
which justifies Eq.~\eqref{eq:dr}. Moreover,
\begin{align}
 \partial_{\bm{k}}\ln\xi=\frac{1-\xi^2}{\xi^2}\frac{\partial_{\bm{k}}\varepsilon}{\varepsilon}
\end{align}
thus the term $\frac12(\bm{s}\times\bm{n})\cdot\dot{\bm{n}}\partial_{\bm{k}}\ln\xi$ has the same direction as the group velocity, it represents the modification of group velocity due to AFM texture.

A further point should be added is that in the most general case, the effective Lagrangian Eq.~\eqref{eq:Lagrangian} should also contain a term representing self-rotation of the wave packet $-\mathrm{Im}[\tilde{c}_i\langle \partial^r_\mu u_i|(\varepsilon-\mathcal{H})|\partial^r_\mu u_j\rangle\tilde{c}_j]$, but after some sophisticated manipulations one can show that this term vanishes for similar reasons as the vanishing $\bm{\Omega}_{\mu\nu}^{kk}$.

\section{}

To study the monopole, we should turn to a different coordinate system. By assigning a variable magnitude to $\bm{n}$, we define the dimensionless order parameter $\bm{R}\equiv R\bm{n}=\frac{J}{t}\bm{n}$. Then the Berry connection can be equivalently defined in the $\bm{R}$ space, which relates to the original one by $A_\mu\mathrm{d}r_\mu=A_i\mathrm{d}R_i$, where
\begin{align}
 A_i&=i \begin{bmatrix}
            \langle\uparrow|\partial_i|\uparrow\rangle& \xi\langle\uparrow|\partial_i|\downarrow\rangle\\
            \xi\langle\downarrow|\partial_i|\uparrow\rangle& \langle\downarrow|\partial_i|\downarrow\rangle
        \end{bmatrix}, \label{eq:monopolepotential}
\end{align}
where $\xi=|\tilde{\gamma}|/\sqrt{R^2+|\tilde{\gamma}|^2}$ is also a function of $R$, and $|\tilde{\gamma}|=\sum_{\bm{\delta}}e^{i\bm{k}\cdot\bm{\delta}}$ depends on the position in BZ.
Written in spherical coordinates, components of Eq.~\eqref{eq:monopolepotential} are
\begin{align}
 A_\mathcal{R}=0,
 A_\theta=\frac{\xi}{2R}
        \begin{bmatrix}
            0 & -i\\
            i & 0
        \end{bmatrix},
 A_\phi=\frac1{2R}
        \begin{bmatrix}
            \cot\theta & -\xi\\
            -\xi & -\cot\theta
        \end{bmatrix}
\end{align}
To see the monopole, we should further make a singular gauge transformation on the potential~\cite{ref:QFT},
\begin{subequations}
\begin{align}
 A_\theta'&=S A_\theta S^\dagger+i\frac1{R} S\partial_\theta S^\dagger \notag\\
          &=i\frac{(1-\xi(R))}{2R}
        \begin{bmatrix}
             0 & e^{-i\phi}\\
            -e^{i\phi} & 0
        \end{bmatrix} \\
 A_\phi'&=S A_\phi S^\dagger+i\frac1{R\sin\theta} S\partial_\phi S^\dagger \notag\\
        &=\frac{(1-\xi(R))}{2R}
        \begin{bmatrix}
             -\sin\theta & e^{-i\phi}\cos\theta\\
             e^{i\phi}\cos\theta & \sin\theta
        \end{bmatrix}
\end{align}
\end{subequations}
with the unitary matrix
\begin{align}
 S=\begin{bmatrix}
  e^{-i\phi/2}\cos\frac\theta2 & -e^{-i\phi/2}\sin\frac\theta2 \\
  e^{i\phi/2}\sin\frac\theta2  & e^{i\phi/2}\cos\frac\theta2
 \end{bmatrix}. \label{eq:Smatrix}
\end{align}
Finally, expressing the gauge potential in Cartesian coordinates, we obtain:
\begin{align}
 A_x&=A_\theta'\cos\theta\cos\phi-A_\phi'\sin\phi=\frac{(1-\xi)}{2R^2}
    \begin{bmatrix}
    y & iz \\
    -iz & -y
    \end{bmatrix} \notag\\
 A_y&=A_\theta'\sin\phi+A_\phi'\cos\phi=\frac{(1-\xi)}{2R^2}
    \begin{bmatrix}
    -x & z \\
    z & x
    \end{bmatrix} \notag\\
 A_z&=A_\theta'\sin\theta=\frac{(1-\xi)}{2R^2}
    \begin{bmatrix}
    0 & -y-ix \\
    -y+ix & 0
    \end{bmatrix} \notag
\end{align}
they are nothing but the ``hedgehog'' gauge potential of a `t~Hooft -Polyakov monopole at $\bm{R}=0$~\cite{ref:QFT},
\begin{align}
 A_i' = \frac{1-\xi(\mathcal{R})}{2R^2}\varepsilon_{ijk}R_j\sigma_k, \label{eq:hedgehog}
\end{align}
where $R_i=\{x,y,z\}$ and $\sigma_i$'s are Pauli matrices. The radial profile of Eq.~\eqref{eq:hedgehog} is determined by the factor $1-\xi(R)$. For fixed nonzero $\tilde{\gamma}$, $1-\xi(R)$ tends to $1$ as $R\rightarrow\infty$, and $1-\xi(R)\sim R^2$ as $R\rightarrow0$, which cancels the $R^2$ in the denominator thus the gauge potential is \emph{regular} at origin.  The form and behaviors of Eq.~\eqref{eq:hedgehog} are all the same as the gauge potential proposed by Sonner and Tong for realizing artificial `t~Hooft-Polyakov monopole~\cite{ref:BPSmonopole}. In fact, our parameter $\xi$ can be understood as the $f(B)$ factor in Ref.~[\onlinecite{ref:BPSmonopole}], they both reflect the overlap of (partial) wave functions from doubly degenerate bands. However, at BZ boundary $\tilde{\gamma}$ is zero, thus $\xi=0$ regardless of $R$. In this case, Eq.~\eqref{eq:hedgehog} becomes the gauge potential of a Wu-Yang monopole and the origin $\bm{R}=0$ becomes singular.

Along with the gauge potential, we can define the associated Higgs field,
\begin{align}
  \phi_H \equiv \begin{bmatrix}
                  \langle A|\sigma_3|A\rangle & \langle A|\sigma_3|B\rangle \\
                  \langle B|\sigma_3|A\rangle & \langle B|\sigma_3|B\rangle
                \end{bmatrix}=\sqrt{1-\xi^2}\ \sigma_3
\end{align}
which is the pseudo-spin polarization of AFM systems. It describes the extent to which the conduction electrons with opposite spins are spatially separated on alternating $A$ and $B$ sites. In other words, it represents how much those electrons respect the staggered order. Upon the same gauge transformation with matrix~\eqref{eq:Smatrix},
\begin{align}
 \phi_H \rightarrow \phi_H'=S\phi_H S^\dagger=\frac{\bm{R}\cdot\bm{\sigma}}{\sqrt{|\tilde{\gamma}|^2+r^2}}=\bm{\Phi}_H\cdot\bm{\sigma}.
\end{align}
The SU(2) gauge field associated with Eq.~\eqref{eq:monopolepotential} is broken into an Abelian magnetic field due to \emph{effective} Higgs mechanism,
\begin{align}
 F_{ij}&=\partial_i(\bm{\Phi}\!\cdot\!\bm{A}_j)-\partial_j(\bm{\Phi}\!\cdot\!\bm{A}_i) +2\bm{\Phi}\cdot(\partial_i\bm{\Phi}\times\partial_j\bm{\Phi}), \\
 B_i&=\frac12\varepsilon_{ijk}F_{jk}=\frac{R_i}{2R^3} \label{eq:DiracMono},
\end{align}
where $\bm{\Phi}=\bm{\Phi}_H/|\bm{\Phi}_H|$, and Eq.~\eqref{eq:DiracMono} is the magnetic field of a Dirac monopole.

An unsolved issue is the Bogomol'nyi relation~\cite{ref:BPSmonopole,ref:QFT}. For the non-Abelian gauge field $\Omega_{ij}=\partial_i A_j-\partial_j A_i-i[A_i,A_j]$, and the covariant derivative $\mathcal{D}_i=\partial_i-i[A_i,\ ]$, it is straightforward to derive
\begin{align}
 \Omega_{ij}=\frac12\varepsilon_{ijk}[\mathcal{D}_k\phi_H-\frac{|\tilde{\gamma}|+\sqrt{|\tilde{\gamma}|^2+R^2}}{(|\tilde{\gamma}|+R^2)^{3/2}}\sigma_k]. \label{eq:Bogomoni}
\end{align}
If not were the last term, Eq.~\eqref{eq:Bogomoni} reproduces the Bogomol'nyi relation. One can show that only for a profile function $\xi(R)=2R/\sinh(R)$ (the case of true 't~Hooft-Polyakov monopole) that the last term vanishes. While our $\xi(R)$ asymptotically resembles $2R/\sinh(R)$, it gives a different profile for finite $R$. As a result, the last term in Eq.~\eqref{eq:Bogomoni} vanishes only when $R\rightarrow\infty$.

\section{}

We eliminate $\mathrm{d}t$ on both sides of Eq.~\eqref{eq:ds},
\begin{align}
 \mathrm{d}\bm{s}=(1-\xi^2)(\bm{s}\cdot\bm{n})\mathrm{d}\bm{n}. \label{eq:dds}
\end{align}
Firstly, dot product both sides of Eq.~\eqref{eq:dds} with $\bm{n}$, regarding the constraint $\bm{n}^2=1$,
\begin{align}
 \bm{n}\cdot\mathrm{d}\bm{s}=(1-\xi^2)(\bm{s}\cdot\bm{n})\frac12\mathrm{d}(\bm{n}^2)=0, \label{eq:nds}
\end{align}
which gives us the relation ${d}(\bm{n}\cdot\bm{s})=\bm{s}\cdot\mathrm{d}\bm{n}$. Secondly, dot product with $\bm{s}$ on both sides of Eq.~\eqref{eq:dds} we get,
\begin{align}
 \mathrm{d}(\bm{s}^2)&=2(1-\xi^2)(\bm{s}\cdot\bm{n})(\bm{s}\cdot\mathrm{d}\bm{n})=(1-\xi^2)\mathrm{d}(\bm{s}\cdot\bm{n})^2 \label{eq:dss}
\end{align}
From Pythagorean theorem we know that $\bm{s}^2=(\bm{s}\cdot\bm{n})^2+(\bm{s}\times\bm{n})^2$, then take derivative on both sides, regarding Eq.~\eqref{eq:dss}, we arrive at
\begin{align}
 \xi^2\mathrm{d}(\bm{s}\cdot\bm{n})^2=-\mathrm{d}(\bm{s}\times\bm{n})^2.
\end{align}
Assume the initial condition $(\bm{s}\cdot\bm{n})|_0=1$, the above equation can be integrated into,
\begin{align}
  (\bm{s}\cdot\bm{n})^2+\frac{(\bm{s}\times\bm{n})^2}{\xi^2}=s_3^2+\frac{s_1^2+s_2^2}{\xi^2}=1,
\end{align}
which justifies Eq.~\eqref{eq:ellipsoid}.


\begin{thebibliography}{20}
  \bibitem{ref:spintronics} I. \v{Z}uti\'{c}, J. Fabian, and S. D. Sarma, Rev. Mod. Phys. \textbf{76}, 323 (2004) and the reference therein.
  \bibitem{ref:BerryPhase} M. V. Berry, Proc. R. Soc. London. A \textbf{392}, 45 (1984).
  \bibitem{ref:Adiabaticity} D. Xiao, M. -C. Zhang, and Q. Niu, Rev. Mod. Phys. \textbf{82}, 1959 (2010) and the reference therein.
  \bibitem{ref:Shengyuan} S. A. Yang \emph{et al.}, Phys. Rev. Lett. \textbf{102}, 067201 (2009); S. A. Yang \emph{et al.}, Phys. Rev. B 82, 054410 (2010).
  \bibitem{ref:Volovik} G. E. Volovik, J. Phys. C \textbf{20}, L83 (1987).
  \bibitem{ref:SMF}  S. E. Barnes and S. Maekawa, Phys. Rev. Lett. \textbf{98}, 246601 (2007).
  \bibitem{ref:Karin} T. Schulz \emph{et al.}, Nature Phys. doi:10.1038/nphys2231 (2012); K. Everschor, M. Garst, R. A. Duine, and A. Rosch, Phys. Rev. B \textbf{84}, 064401 (2011).
  \bibitem{ref:THE} M. Lee, W. Kang, Y. Onose, Y. Tokura, and N. P. Ong, Phys. Rev. Lett. \textbf{102}, 186601(2009); A. Neubauer \emph{et al.}, Phys. Rev. Lett. \textbf{102}, 186602 (2009); P. Bruno, V. K. Dugaev, and M. Taillefumier, Phys. Rev. Lett. \textbf{93}, 096806 (2004); J. Ye \emph{et al.}, Phys. Rev. Lett. \textbf{83}, 3737 (1999).
  \bibitem{ref:ZhangShoucheng} Y. B. Bazaliy, B. A. Jones, and S. -C. Zhang, Phys. Rev. B \textbf{57}, R3213 (1998).
  \bibitem{ref:Tserkovnyak} C. H. Wong and Y. Tserkovnyak, Phys. Rev. B \textbf{80}, 184411 (2009); Y. Tserkovnyak and C. H. Wong, Phys. Rev. B \textbf{79}, 014402 (2009).
  \bibitem{ref:STT} J. Zang, M. Mostovoy, J. H. Han, and N. Nagaosa, Phys. Rev. Lett. \textbf{107}, 136804 (2011); F. Jonietz \emph{et al.}, Science \textbf{330}, 1648 (2010).
  \bibitem{ref:AFMTheory} A. C. Swaving and R. A. Duine, J. Phys.: Cond. Mat. \textbf{24}, 024223 (2012); K. M. D. Hals, Y. Tserkovnyak, and A. Brataas, Phys. Rev. Lett. \textbf{106}, 107206 (2011); R. Wieser, E. Y. Vedmedenko, and R.~Wiesendanger, Phys. Rev. Lett. \textbf{106}, 067204 (2011); A. C. Swaving and R. A. Duine, Phys. Rev. B \textbf{83}, 054428 (2011); Y. Xu, S. Wang, and K. Xia, Phys. Rev. Lett. \textbf{100}, 226602 (2008); P. M. Haney and A. H. MacDonald, Phys. Rev. Lett. \textbf{100}, 196801 (2008).
  \bibitem{ref:AFMExperiment} S. Urazhdin and N. Anthony, Phys. Rev. Lett. \textbf{99}, 046602 (2007); R. Jaramillo \emph{et al.}, Phys. Rev. Lett. \textbf{98}, 117206 (2007); Z. Wei \emph{et al.}, Phys. Rev. Lett. \textbf{98}, 116603 (2007).
  \bibitem{ref:AFMSpintronics} A. H. MacDonald and M. Tsoi, Phil. Trans. R. Soc. A \textbf{369}, 3098 (2011).
  \bibitem{ref:Dimi} D. Culcer, Y. Yao, and Q. Niu, Phys. Rev. B \textbf{72}, 085110 (2005).
  \bibitem{ref:Dalibard} J. Dalibard, \emph{et al.}, Rev. Mod. Phys. \textbf{83}, 1523 (2011).
  \bibitem{ref:NABerryPhase} F. Wilczek and A. Zee, Phys. Rev. Lett. \textbf{52}, 2111 (1984); J. Moody, A. Shapere, and F. Wilczek, Phys. Rev. Lett. \textbf{56}, 893 (1986); C. A. Mead, Phys. Rev. Lett. \textbf{59}, 161 (1987); B. Zygelman, Phys. Rev. Lett. \textbf{64}, 256 (1990).
  \bibitem{ref:BPSmonopole} J. Sonner and D. Tong, Phys. Rev. Lett. \textbf{102}, 191801 (2009).
  \bibitem{ref:SpiralingSpin} F. Y. Yang and C. L. Chien, Phys. Rev. Lett. \textbf{85}, 2597 (2000).
  \bibitem{ref:private} J. -W. Cai, \textit{private communications}.
  \bibitem{ref:Mnlayer} M. Bode \textit{et al.}, Nature \textbf{447}, 190 (2007).
  \bibitem{ref:noteisospin} Different from the peudospin furnished by the $A-B$ sublattices, isospin refers to the superposition coefficients $c_a$ and $c_b$. Only in the limit $\xi\rightarrow0$ where $|A\rangle$ and $|B\rangle$ are orthogonal, isospin and peudospin become equivalent.
  \bibitem{ref:QFT} Ya. Shnir, \emph{Magnetic Monopoles}, Springer-Verlag Berlin Heidelberg, 2005.
\end{thebibliography}
\end{document}